\documentclass[ %
 reprint,
superscriptaddress,
 amsmath,amssymb,
 aps,
prb,
]{revtex4-1}

\usepackage{graphicx}
\usepackage{dcolumn}
\usepackage{bm}
\usepackage{subfigure}
\usepackage{epstopdf}
\usepackage{lettrine}

\begin{document}

\preprint{APS/123-QED}

\title{Interference Effects in a Tunable QPC-Cavity-Reflector Integrated System}

\author{Chengyu Yan}
 \email{uceeya3@ucl.ac.uk}
 \affiliation{%
 London Centre for Nanotechnology, 17-19 Gordon Street, London WC1H 0AH, United Kingdom\\
 }%
 \affiliation{
  Department of Electronic and Electrical Engineering, University College London, Torrington Place, London WC1E 7JE, United Kingdom
 }%
 \author{Sanjeev Kumar}
 \affiliation{%
 London Centre for Nanotechnology, 17-19 Gordon Street, London WC1H 0AH, United Kingdom\\
 }%
 \affiliation{
  Department of Electronic and Electrical Engineering, University College London, Torrington Place, London WC1E 7JE, United Kingdom
 }%
\author{Michael Pepper}
\affiliation{%
 London Centre for Nanotechnology, 17-19 Gordon Street, London WC1H 0AH, United Kingdom\\
 }%
 \affiliation{
  Department of Electronic and Electrical Engineering, University College London, Torrington Place, London WC1E 7JE, United Kingdom
 }%
\author{Patrick See}
\affiliation{%
 National Physical Laboratory, Hampton Road, Teddington, Middlesex TW11 0LW, United Kingdom\\
}%
\author{Ian Farrer}
\email{ Department of Electronic and Electrical Engineering, University of Sheffield, Mappin Street, Sheffield S1 3JD, United Kingdom\\}
\affiliation{%
  Cavendish Laboratory, J.J. Thomson Avenue, Cambridge CB3 OHE, United Kingdom\\
}%
\author{David Ritchie}
\affiliation{%
 Cavendish Laboratory, J.J. Thomson Avenue, Cambridge CB3 OHE, United Kingdom\\
}%
\author{Jonathan Griffiths}
\affiliation{%
 Cavendish Laboratory, J.J. Thomson Avenue, Cambridge CB3 OHE, United Kingdom\\
}%
\author{Geraint Jones}
\affiliation{%
 Cavendish Laboratory, J.J. Thomson Avenue, Cambridge CB3 OHE, United Kingdom\\
}%

\date{\today}

\begin{abstract}
We show experimentally how quantum interference can be produced using an integrated quantum system comprising an arch-shaped short quantum wire (or quantum point contact, QPC) of 1D electrons and a reflector forming an electronic cavity. On tuning the coupling between the QPC and the electronic cavity, fine oscillations are observed when the arch-QPC is operated in the quasi-1D regime. These oscillations correspond to interference between the 1D states and a state which is similar to Fabry-Perot and suppressed by a small transverse magnetic field of $\pm$60 mT. Tuning the reflector we find a peak in resistance which follows the behavior expected for a Fano resonance. We suggest that this is an interesting example of a Fano resonance in an open system which corresponds to interference at or near the Ohmic contacts due to a directly propagating, reflected discrete path and the continuum states of the cavity corresponding to multiple scattering. Remarkably, the Fano factor shows an oscillatory behavior taking peaks for each fine oscillation thus confirming coupling between the discrete and continuum states. The results indicate that such simple quantum device could be used  as building blocks to create more complex integrated quantum circuits for possible applications ranging from quantum information processing to realizing the fundamentals of complex quantum systems.

\end{abstract}

\maketitle

\section{Introduction}

Quantum interference, one of the most remarkable effects of quantum mechanics, arising from the wave nature of particles has led to some celebrated results in mescoscopic system, including weak localization in two dimensional (2D) systems\cite{MRM81,RMM81}, Aharonov-Bohm oscillations in ring structures\cite{CTR88,BAS99}, sharp peaks in magnetoresistance in chaotic cavity\cite{TND95,BDN95} , etc. Among its various applications, quantum interference has been used successfully as a tool to investigate properties of particles such as monitor correlation and entanglement as demonstrated in Mach-Zehnder interferometery\cite{JCS03,BVD06}, Aharonov-Bohm interferometery\cite{NOC07,HDQ01}, Hanbury Brown-Twiss interferometery\cite{JJW07,KRA02}, and electronic analogue of Hong-Ou-Mandel device\cite{LOY98,BFB13}. Among all the striking phenomenon, it is particularly interesting to note that quantum interference is extremely suitable for studying coupling between different quantum systems\cite{MTM94,CDO15,JMY14} which is crucial for the design of integrated quantum circuits. 

\begin{figure}

	     \subfigure{
		\includegraphics[height=2.4in,width=2.8in]{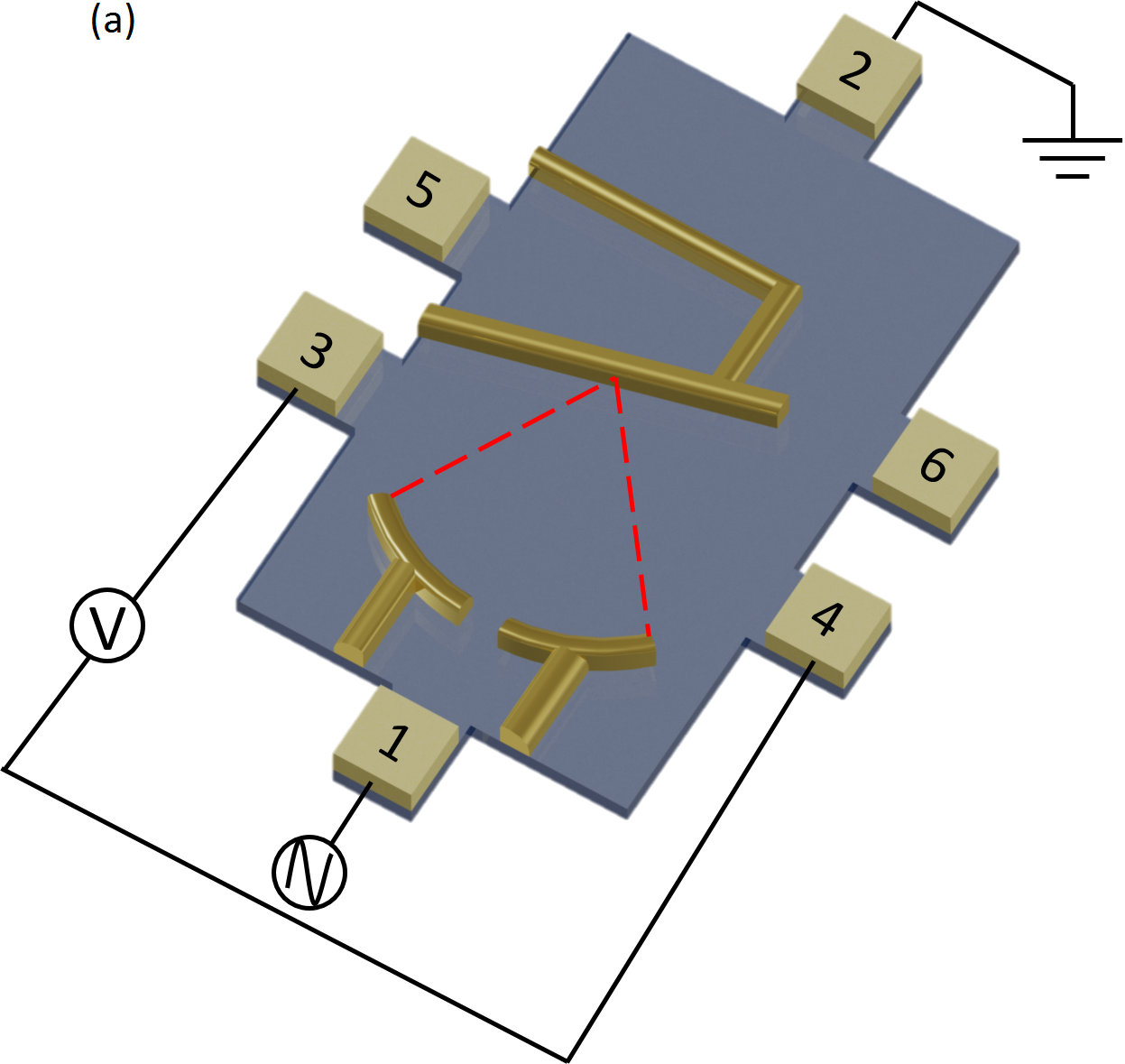}
		\label{fig:Fig1a_YAN}
	}%
	
	\subfigure{
		\includegraphics[height=2.0in,width=3.2in]{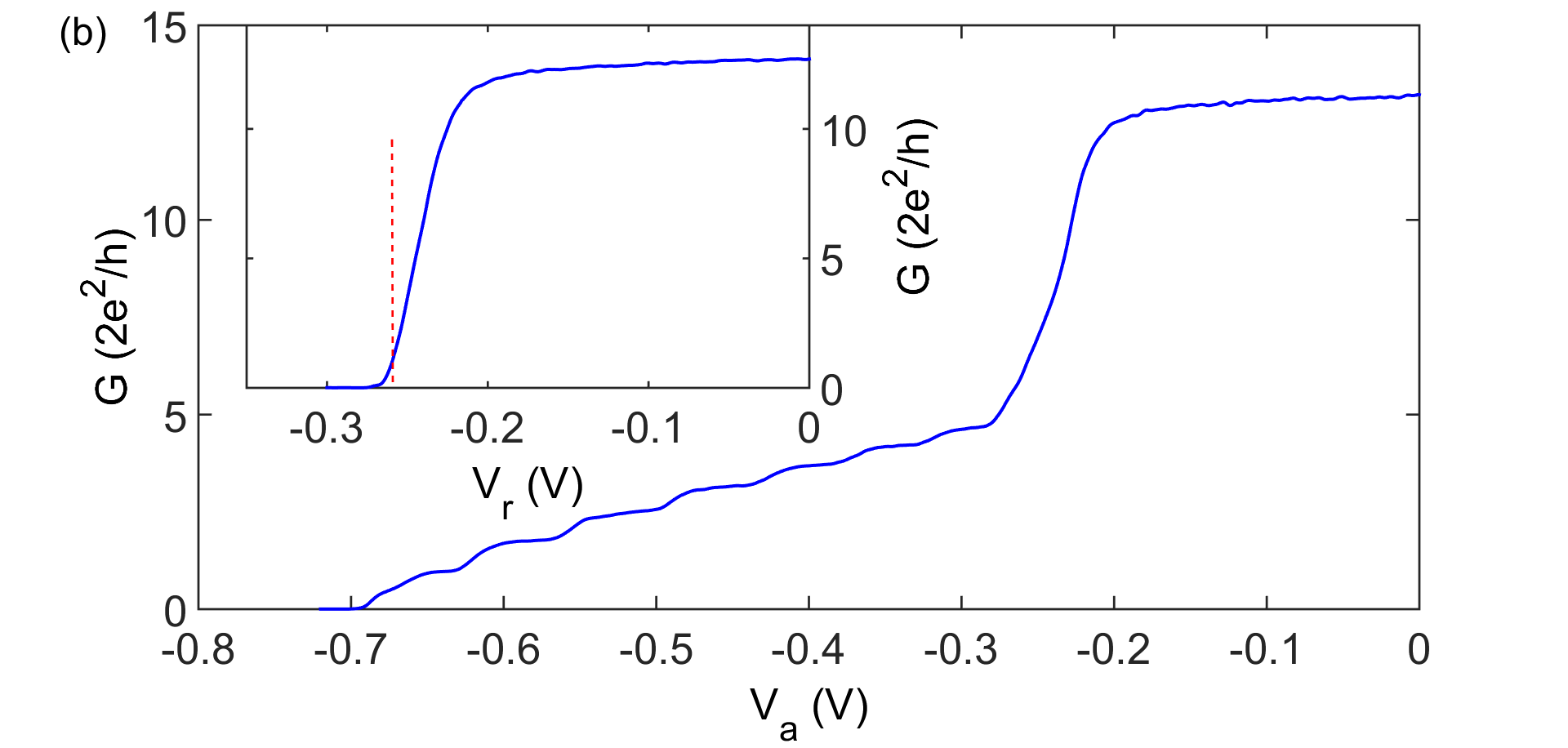}
		\label{fig:Fig1b_YAN}
	}%
	
	\caption{Schematic of device and experiment setup. (a) The yellow-square-blocks at the edge of mesa marked 1-6 are Ohmics contacts whereas the golden-metallic patterns within the mesa form a pair of arch-shaped gates and a reflector. The red dotted-lines show the possible boundary of cavity formed between the QPC and the reflector. Excitation current is fed to Ohmic 1 while 2 is grounded; Ohmics 3 and 4 are voltage probes. (b) Differential conductance measurement of the QPC (main plot) and reflector (inset). The series resistance has not been removed. It should be noticed that the standard two-terminal measurement is performed with Ohmics 1 and 5\cite{note1}. The red-dashed line in the inset corresponds to $V_r$ = - 0.27 V.  }           
	\label{fig:exp_setup}
\end{figure}

A notable quantum mechanical system is a ballistic quantum wire, when short called a quantum point contact (QPC), defined using a pair of split gates\cite{TPA86,DTM88,WHB88} on a semiconductor heterostructure. Soon after the first observation of quantised conductance in a quasi-one-dimensional (1D) quantum wire\cite{DTM88,WHB88}, a number of novel interesting observations have been made such as the `0.7 structure'\cite{KNM96}, incipient Wigner lattice\cite{KTM14,HTM09}, and coherent electron focusing\cite{CHB88,HVH89,CPF12}, etc. Experiments regarding coupling between a QPC, which provides a stream of collimated 1D electrons, and waveguide\cite{MTM94} and that between quantum dot (QD) and cavity\cite{CDO15} have been performed, where a pronounced modulation of conductance of the system and changing of Coulomb peaks were observed. Knowledge on the resonant interference effects arising from the coupling of two quantum states is of fundamental significance for the development of technology for quantum information processing. Moreover, the Fano resonance, which arises from the quantum interference between discrete (bound) and continuum states of a neighboring medium resulting in an asymmetry in the resonance structure\cite{FANO61,PP15}, is one of the definite methods for confirming quantum interference particularly in systems with smaller length scales, and tunable electron path. We present a system as shown in Fig.~\ref{fig:exp_setup} realized using an integrated quantum circuit having three components: a source of 1D electrons (emitted by a QPC), a continuum state (an electronic cavity of 2D electrons) and a reflector (a potential barrier to reflect the collimated electrons). Such an ensemble would allow a beam of 1D electrons to interfere with the continuum states of the cavity after being reflected by the potential barrier. This system was used in our previous work\cite{CSM17} where we showed that a Fano resonance could be observed at the 1D-2D transition regime of a QPC. In the present work we give evidence that a bound, localized state can be established within the continuum (2D) and there is a coupling between the bound state and the 1D states from which it is drawn. Moreover, we show that the coupling between the 1D-2D states enables the Fano resonance which can be tuned both electrostatically and magnetically. 

The paper from now onwards is organized in the following manner: Section II (Experimental) covers the experimental setup, device design, and wafer details. Section III (Results and Discussion) covers details of comprehensive experimental results. This section is further divided into four sub-sections: A. Coupling between the 1D-2D electrons; B. Fano resonance; C. Effect of perpendicular magnetic field; and D. Temperature dependence.   

\section{Experimental}

\begin{figure}	  
\includegraphics[height=2.0in,width=3.2in]{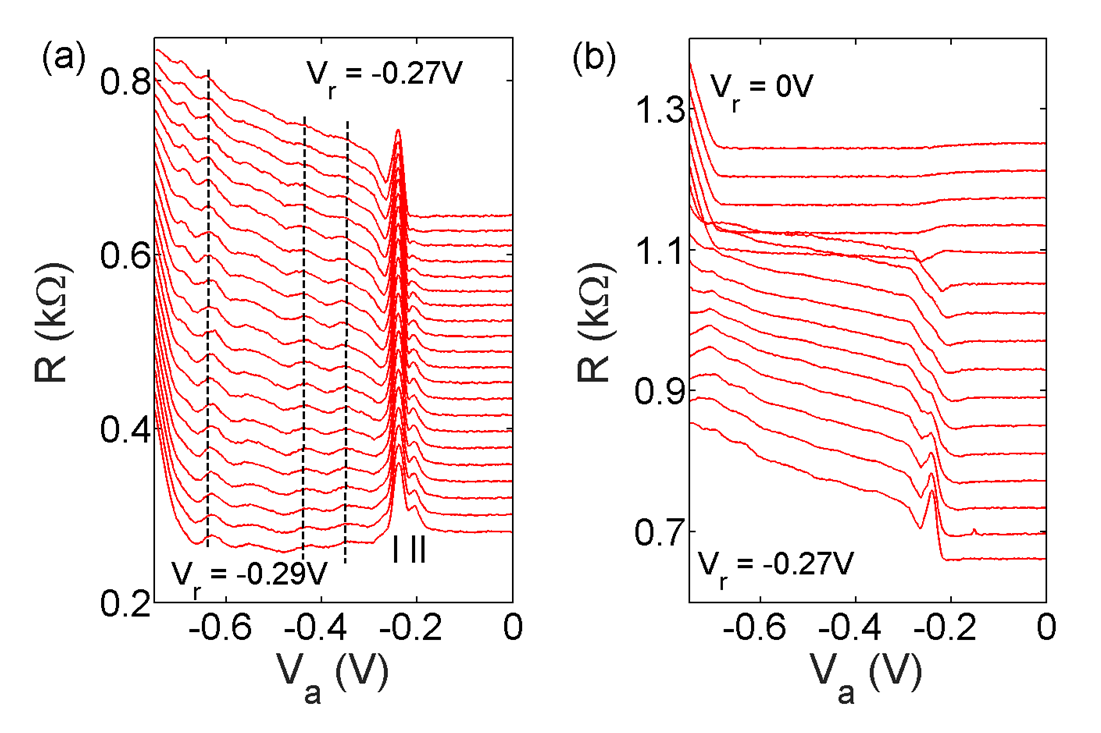} 
\caption{\textit{R} as a function of arch-QPC voltage for various V$_r$. (a) When -0.29 V $\leq V_r \leq$ -0.27 V, a pronounced double-peak structure and fine oscillations are observed. (b) When $V_r >$ -  0.27 V, all the features gradually smear out. Data have been offset vertically by 20 $\Omega$ for clarity.  } 
\label{fig:QPC_ref_together}
\end{figure}

The samples consist of a pair of arch-shaped gates, with a QPC forming in the center of the arch, and a reflector inclined at 75$^\circ$ to the current flow direction such that the center of the QPC and the reflector are aligned  as shown in Fig.~\ref{fig:exp_setup}(a). The opening angle of the arch is 45$^\circ$ and the radius is 1.5 $\mu$m, both the length and width of the QPC embedded in the arch is 200 nm, the width of the reflector is 300 nm\cite{CSM17}. The angle of the inclined reflector is crucial because it allows reflected electrons to propagate ballistically to Ohmic contact 3, while multiple scattering of others will set up a continuum of cavity states (considering the normal 30$^\circ$ spread of the collimated 1D electrons\cite{LAC90}). A direct scattering of a beam of electrons from the reflector at or near the entrance of the QPC will establish the bound, defined, state. The scattering of electrons out of this state  interfering with cavity states will establish a Fano resonance. 

The devices were  fabricated on a high mobility two-dimensional electron gas formed at the interface of GaAs/Al$_{0.33}$Ga$_{0.67}$As heterostructure. The electron density (mobility) measured at 1.5 K was 1.80$\times$10$^{11}$cm$^{-2}$ (2.1$\times$10$^6$cm$^2$V$^{-1}$s$^{-1}$) therefore both mean free path and phase coherence length were over 10 $\mu$m which is much larger than the distance between the QPC and reflector (1.5 $\mu$m). All the measurements were performed with the standard lock-in technique in a cryofree dilution refrigerator with a lattice temperature of 20 mK. For the four-terminal resistance measurement a 10 nA at 77 Hz ac current is applied while an ac voltage of 10 $\mu$V at 77 Hz is used for the two-terminal conductance measurement. Figure~\ref{fig:exp_setup}(b) shows the conductance trace of the QPC with well defined conductance plateaus. On the other hand, the conductance of the reflector (inset) drops around - 0.2 V which indicates a sharp change in the transmission probability; here V$_a$ and V$_r$ are the voltages applied on the arch gates forming the QPC and reflector, respectively.

\section{Results and discussion}

When the arch-QPC and reflector are operated together, collimated ballistic 1D electrons\cite{LAC90} injected from the QPC are reflected by the potential barrier created by the reflector and cause a voltage drop $V_{34}$ between Ohmics 3 and 4, such that $$V_{34} \propto n \times r \eqno(1)$$ where \textit{n} is  the population of injected electrons and \textit{r} is the reflection probability\cite{CSM17}.

Figure~\ref{fig:QPC_ref_together} shows the four-terminal resistance measured between Ohmics 3 and 4 as a function of arch-QPC voltage for various reflector voltages V$_r$. It is found that when - 0.29 V $\leq V_r \leq$ - 0.27 V, a pronounced double-peak structure is observed when arch-QPC voltage V$_a$ is around - 0.2 V, the left peak at V$_a$ = - 0.24 V is referred to as peak I and the right peak at - 0.21 V as peak II [Fig.~\ref{fig:QPC_ref_together}(a)]. When making $V_r$ less negative, peak II weakens gradually while the peak I remains robust. Interestingly, the position of the double-peak structure coincides with the 1D-2D transition regime of the QPC. In addition, fine oscillations, highlighted by the vertical dashed lines, occur when the arch-QPC forms a quasi-1D channel. When $V_r >$ - 0.27 V, both the double-peak structure and fine oscillations weaken and eventually smear out in this regime [Fig.~\ref{fig:QPC_ref_together}(b)].

It is interesting to note that the fine oscillations almost align with conductance plateaus in a similar manner to that reported in tunnelling spectroscopy of a waveguide\cite{MTM94,CJ91}. However, unlike the results in waveguide experiment\cite{MTM94,CJ91} where each subband contributes to a sharp peak, conductance plateaus at - 0.6 and - 0.5 V correlate to a broad structure or oscillation in our work. The double-peak structure in the 1D-2D transition regime has no analogue in the waveguide structure\cite{CJ91,YDK92,YD92}. The correlation between fine oscillations and 1D density of states (DOS) cannot explain the strong sensitivity of the additional structures on tuning the reflector voltage.

We can visualize our system as consisting of a dynamic electronic cavity\cite{CDO15} which is defined by the arch-gate and the reflector (the reflector is required to define the focal point of the cavity). The multiple reflection of emitted 1D electrons between the reflector and the arch-gates give rise to a continuum of localized states. When the electrons under the gates are depleted and thus the cavity is switched on, the 1D states couple to the cavity states and thus give rise to the observed non-trivial features. 

The framework above can be verified by changing the coupling between the 1D and cavity states. A strong coupling will result in the appearance of fine oscillations while at weak coupling they will smear out.

\subsection{Coupling between 1D-2D electrons}

\begin{figure}
     \subfigure{
		\includegraphics[height=2.0in,width=3.2in]{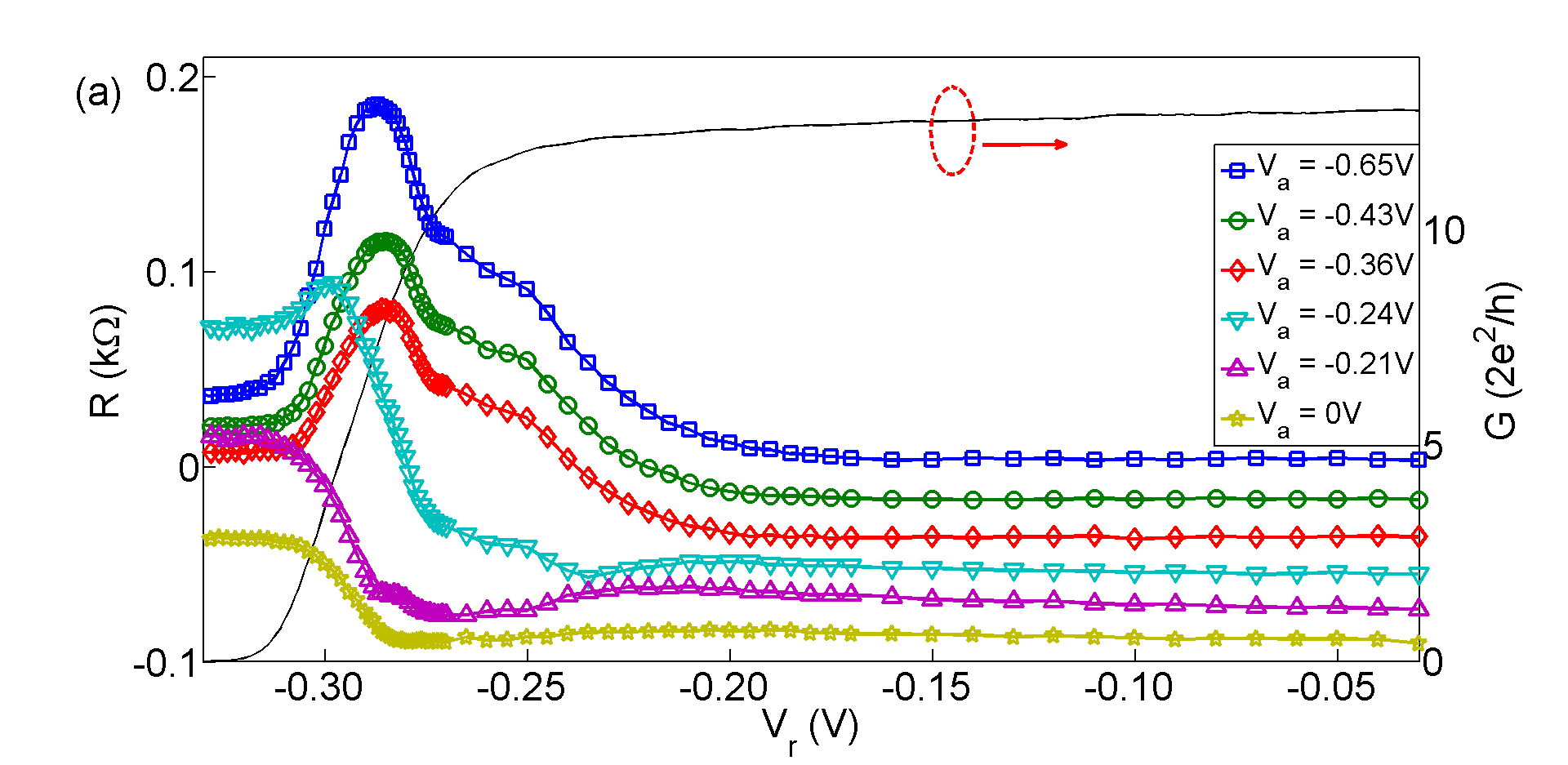}
		\label{fig:measurement_reflector}
	}%
	
	\subfigure{
		\includegraphics[height=1.8in,width=3.5in]{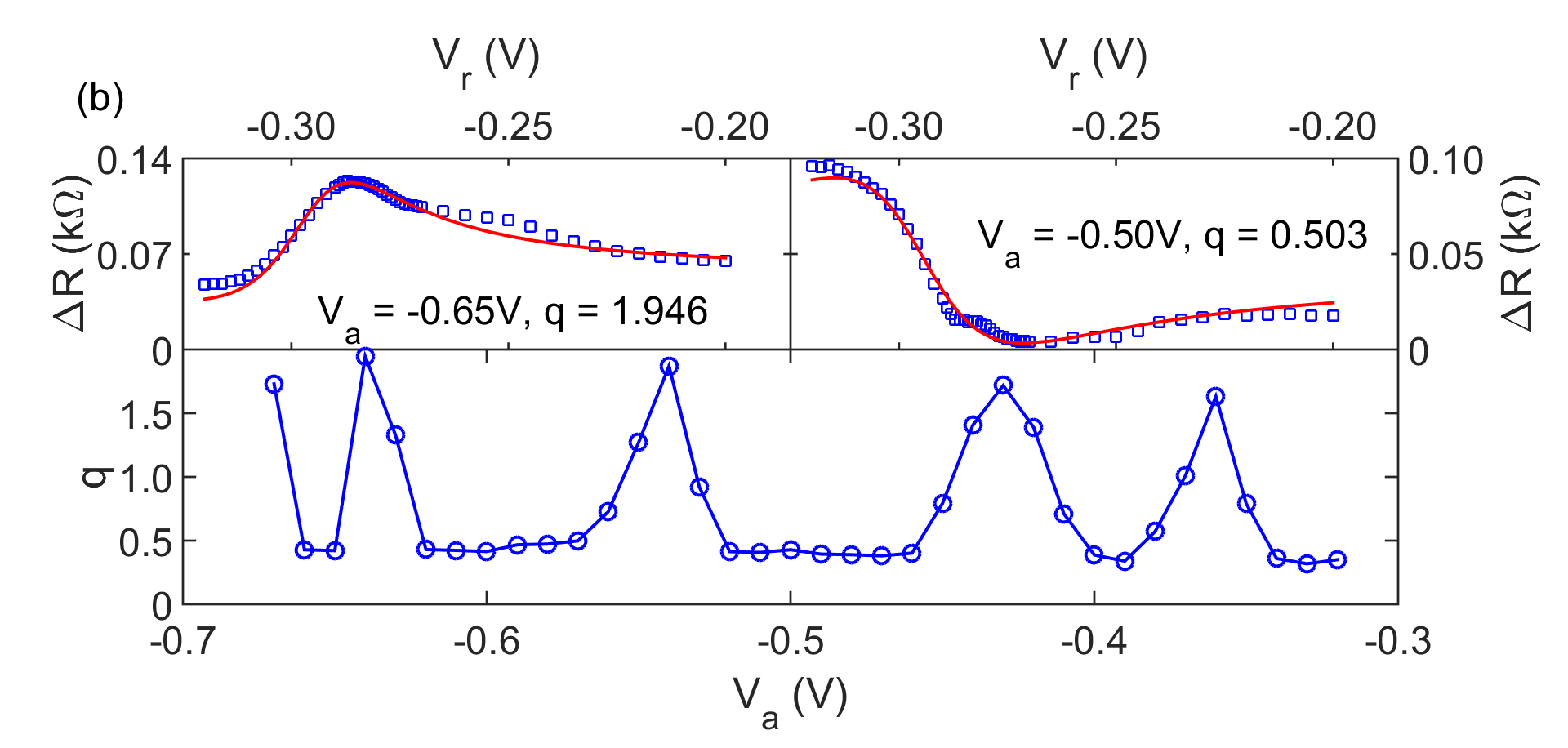}
		\label{fig:Fano_reflector}
	}%
	\caption{\textit{R} as a function of V$_r$. (a) The arch-gate voltage is fixed at - 0.65 , - 0.43, - 0.36, - 0.24, - 0.21, and 0 V while sweeping V$_r$. Data have been offset vertically for clarity. The right axis displays the reflector conductance as a function of V$_r$ as indicated by a red arrow. (b) Upper plots show the theoretical fitting (solid line) of $\Delta$R(V$_r$,V$_a$) = R(V$_r$,V$_a$) - R(V$_r$,0), where V$_a$ = - 0.65 and - 0.50 V, respectively; $\Delta$R follows a well defined Fano line shape. Lower plot shows Fano factor \textit{q} as a function of V$_a$.}           
	\label{fig:reflector_sweep}
\end{figure}

 To tune the coupling between the 1D and 2D cavity states, we fix the arch-QPC voltage at  -0.65, - 0.43, - 0.36 [these three voltages are the center of fine oscillations, as indicated by the vertical lines in Fig.~\ref{fig:QPC_ref_together}(a)], - 0.24 (Peak I), - 0.21 (Peak II) and 0 V and sweep the reflector voltage. The result shown in Fig.~\ref{fig:reflector_sweep}(a) is a direct measurement of strength of the coupling effect. When the arch-QPC voltage is fixed the population of injected electrons \textit{n} in Eq.(1) is constant, so the measured resistance depends on the reflection probability \textit{r} only; \textit{r} increases monotonically when V$_r$ becomes more negative, therefore resistance \textit{R} should follow a similar monotonic change. The bottom two traces in Fig.~\ref{fig:reflector_sweep}(a) are for arch-QPC voltages V$_a$ of 0 and - 0.21 V, respectively (cavity is off). The resistance is initially almost constant when the reflector voltage $V_r >$ - 0.25 V. A sharp rise in resistance occurs as V$_r$ is made more negative so the reflection probability \textit{r} increases rapidly, and eventually saturates when \textit{r} becomes unity. When the arch-QPC becomes more negative (top four traces, the cavity turns on), $R$ initially keeps on increasing and then decreases until it saturates, and thereby produces a pronounced peak in the plot. It is necessary to emphasize that the reflector voltage range (from - 0.27 to - 0.29 V) where the peak in resistance occurs corresponds to the regime where the fine oscillations are most pronounced [Fig.~\ref{fig:QPC_ref_together}(a)]. The dramatic change between different traces is a clear indication of the effect arising due to cavity formation. Once the cavity is on, i.e. both V$_a$ and V$_r$ are sufficiently negative, its size and therefore the energy spacing of cavity states can be adjusted by sweeping the reflector voltage. The strong coupling between the 1D and cavity leads to a peak in the measurement.
 
 The fine oscillations corresponding to interference between the waves emerging from the 1D region and the reflected waves are observable as the phase coherence length ($>$10 $\mu$m) is greater than the size of the device. In this respect our observation is similar to the Fabry-Perot interference previously observed\cite{CMH89}.

\subsection{Fano resonance} 

An analysis of the line shape of peak in Fig.~\ref{fig:reflector_sweep}(a) clearly indicates that after the cavity is switched on, the dynamic change in resistance $\Delta$R(V$_r$, V$_a$) = R(V$_r$, V$_a$) - R(V$_r$, 0) (we use R(V$_r$, 0) to account for the change in the reflection probability, which is always present) follows the well known Fano resonance\cite{FANO61,GGH00,CSM17}, $$ R = R_0\frac{(q + \gamma (V - V_0))^2}{1 + \gamma^2(V - V_0)^2} + R_{inc}\eqno(2)$$ where R is the measured resistance, R$_0$ is a constant representing the amplitude of the resonance, q is the Fano factor which decides the asymmetry of the line shape, $\gamma$ = 20 V$^{-1}$ is voltage-energy lever (estimated from the Fermi energy and the pinch-off voltage), V$_0$ is the arch-gate voltage at the center of the resonance (dip), and R$_{inc}$ denotes intrinsic contribution from the background  as shown in Fig.~\ref{fig:reflector_sweep}(b). It was noticed that the Fano factor \textit{q} takes an oscillatory behavior against the arch-QPC voltage as shown in the lower graph in Fig.~\ref{fig:reflector_sweep}(b). When compared with Fig.~\ref{fig:QPC_ref_together}, \textit{q} has a peak ($\sim$ 1.9) at the fine oscillations and remains constant ($\sim$ 0.4) elsewhere. The increase in \textit{q} is due to an increase in the resonant scattering.  This result supports the argument that the observed fine oscillations are correlated with the Fano resonance. It is necessary to draw attention to the fact that the Fano factor does not necessarily show a peak at a conductance plateau if there is no fine oscillation as two conductance plateaus are well defined when - 0.6 V $< V_a <$ - 0.5 V while only a single Fano factor peak is observed. This suggests that the result does not arise simply due to the collimation of a QPC (which will produce similar behavior at all the conductance plateaus) or an electrostatic effect.

The energy spacing between different cavity states is small (of the order of 100 $\mu$eV, from Ref.16)  due to the large size of the cavity. This indicates one 1D-state may couple to several cavity states and could result in relatively broad fine oscillations. On the other hand, a single broad pattern from 2$^{nd}$ (V$_a$ = - 0.6 V) and 3$^{rd}$ (V$_a$ = - 0.5 V) plateaus or modes of the QPC may be attributed to coupling of different modes of QPC with the degenerate cavity states. In addition, the probability of intersubband transition in 1D constriction\cite{OFM93} may result in mixed states.

Most investigations of Fano Resonance in nanostructures have utilized a discrete state, for example in a quantum dot\cite{GGH00} or coupled QPCs\cite{JMY14}. It is clear that the system which has been studied here has significant differences in the lack of control of the particular states undergoing interference as well as being open\cite{RLB04,SJ05,MY05}. However, the discrete state, forming part of the interference  along with the continuum, corresponds to a collimated electron wave between QPC and cavity that is then reflected towards the Ohmic contact and subsequently interferes. This is essentially identical to the case of an optical wave reflected by a sharp bend in a waveguide which behaves as a localized state\cite{MTM94,CJ91}. The momentum value and direction required for the reflected wave to reach the contact without further reflection are stringent, corresponding to a sharply defined state which then can interfere with the continuum comprising scattered electrons.

\subsection{Effect of perpendicular magnetic field}

\begin{figure}

	\includegraphics[height=2.0in,width=3.6in]{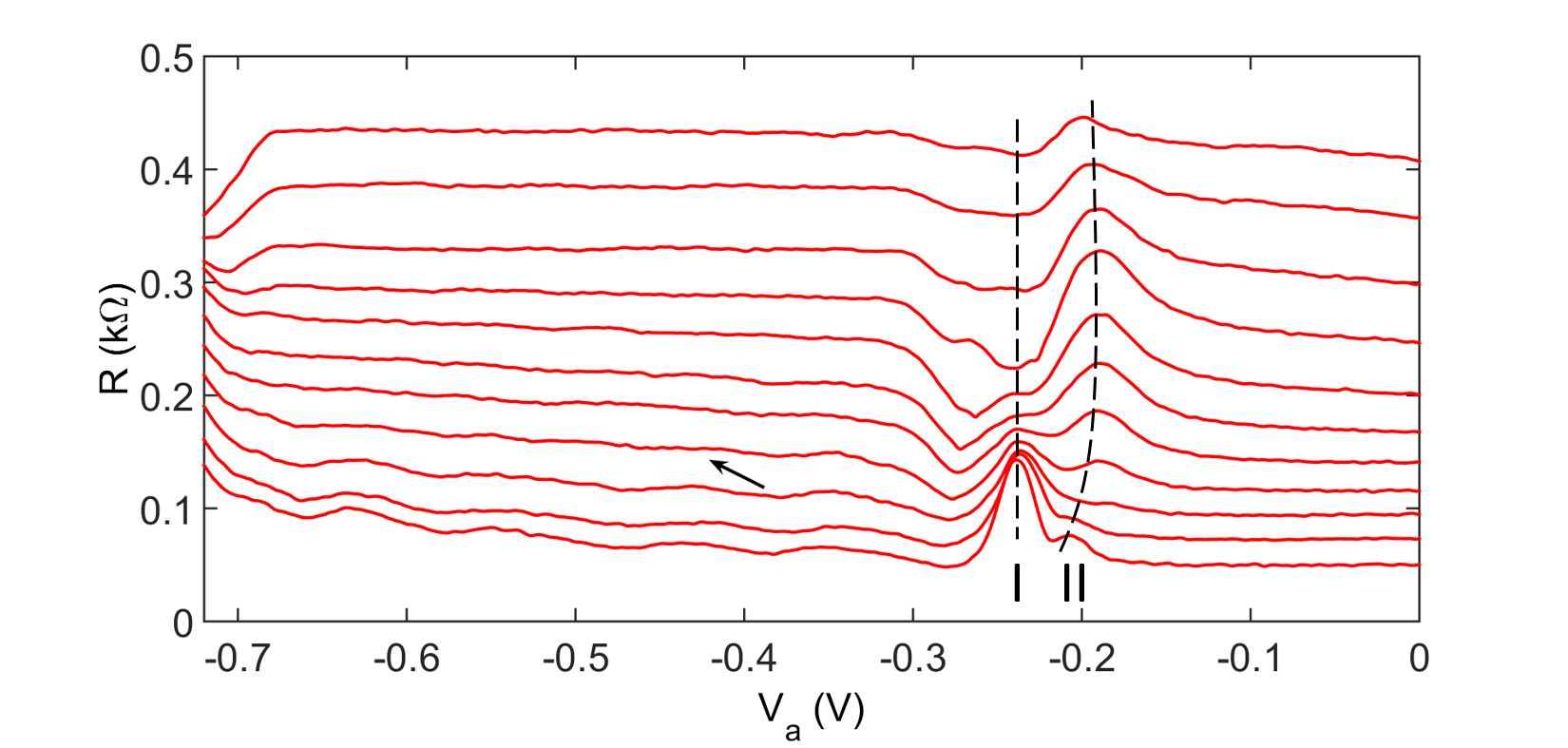}

	\caption{Effect of perpendicular magnetic field. The reflector voltage is set to -0.28 V and the field is increased from 0 (bottom trace) to - 200 mT (top trace) by steps of - 20 mT. It  is clear that the fine oscillations disappear around - 60 mT (marked by an arrow), peak I of the double-peak structure weakens with increasing field while peak II gets enhanced by magnetic field and shifts towards less negative V$_{a}$ as indicated by a dashed black curve. 
	}           
	\label{fig:field_raw_data}
\end{figure}

\begin{figure}
	\subfigure{
		\includegraphics[height=1.8in,width=3.5in]{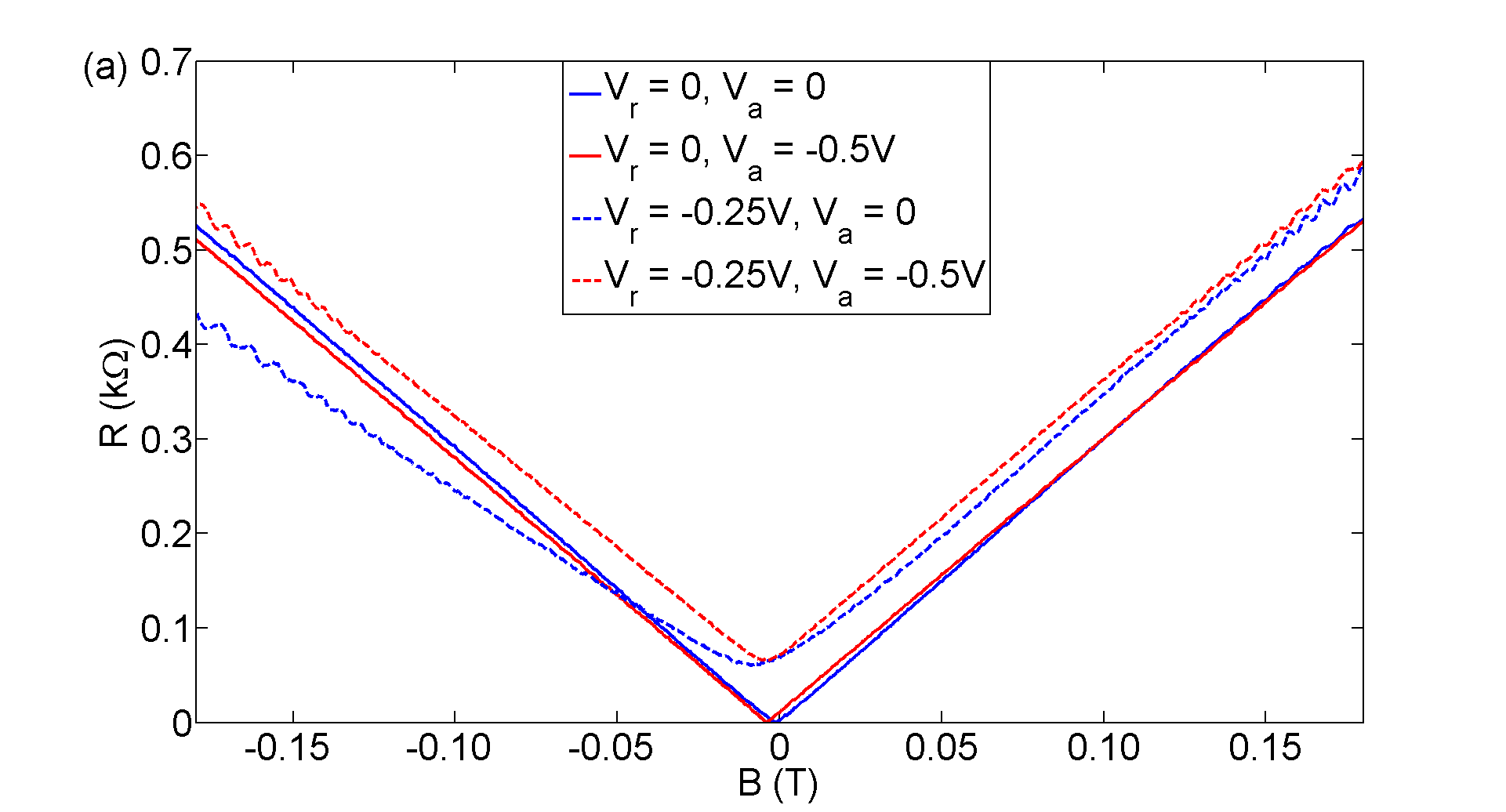}
		\label{fig:B_dep_off}
	}%
	
	\subfigure{
		\includegraphics[height=1.8in,width=3.5in]{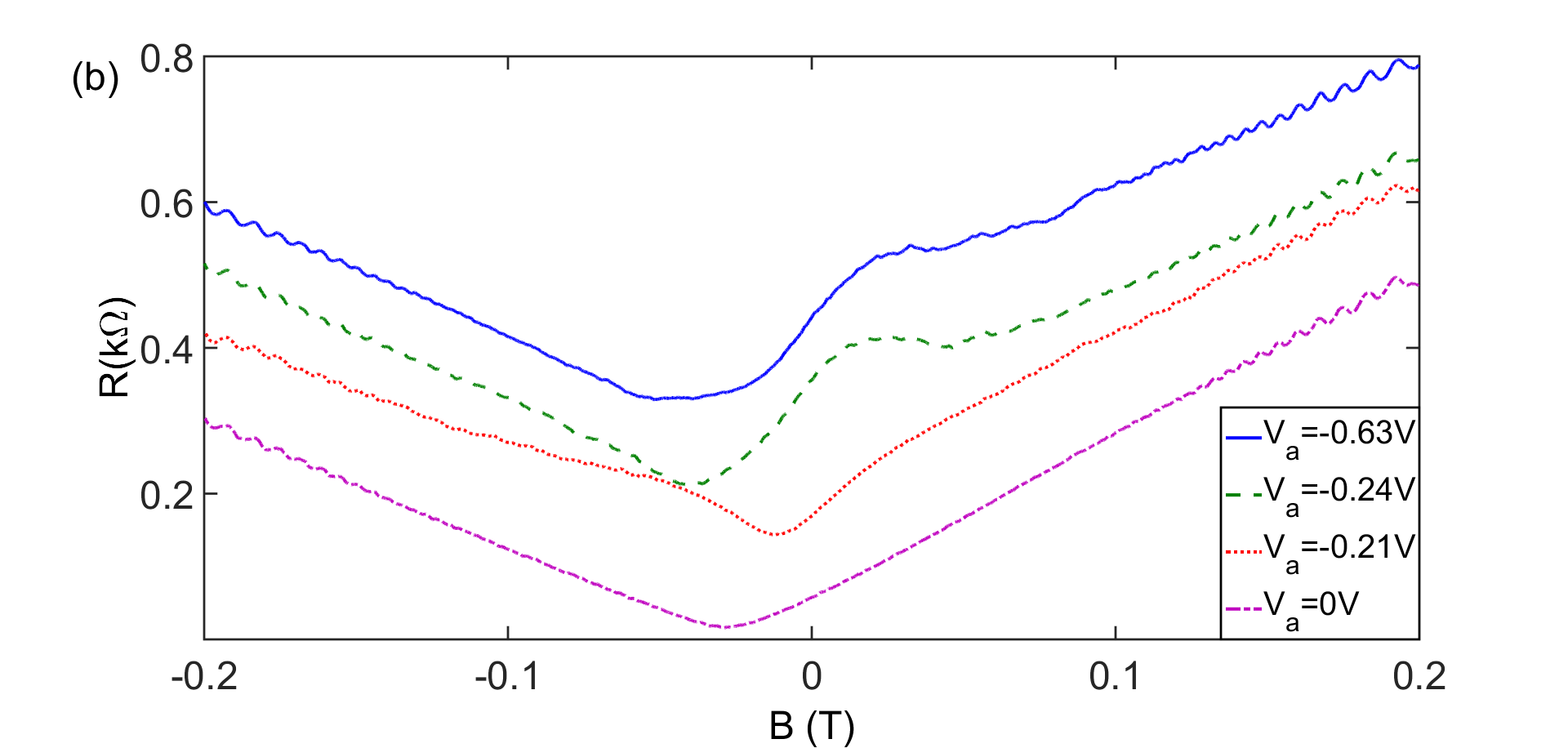}
		\label{fig:B_dep_on}
	}%
	
	\subfigure{
		\includegraphics[height=1.8in,width=3.5in]{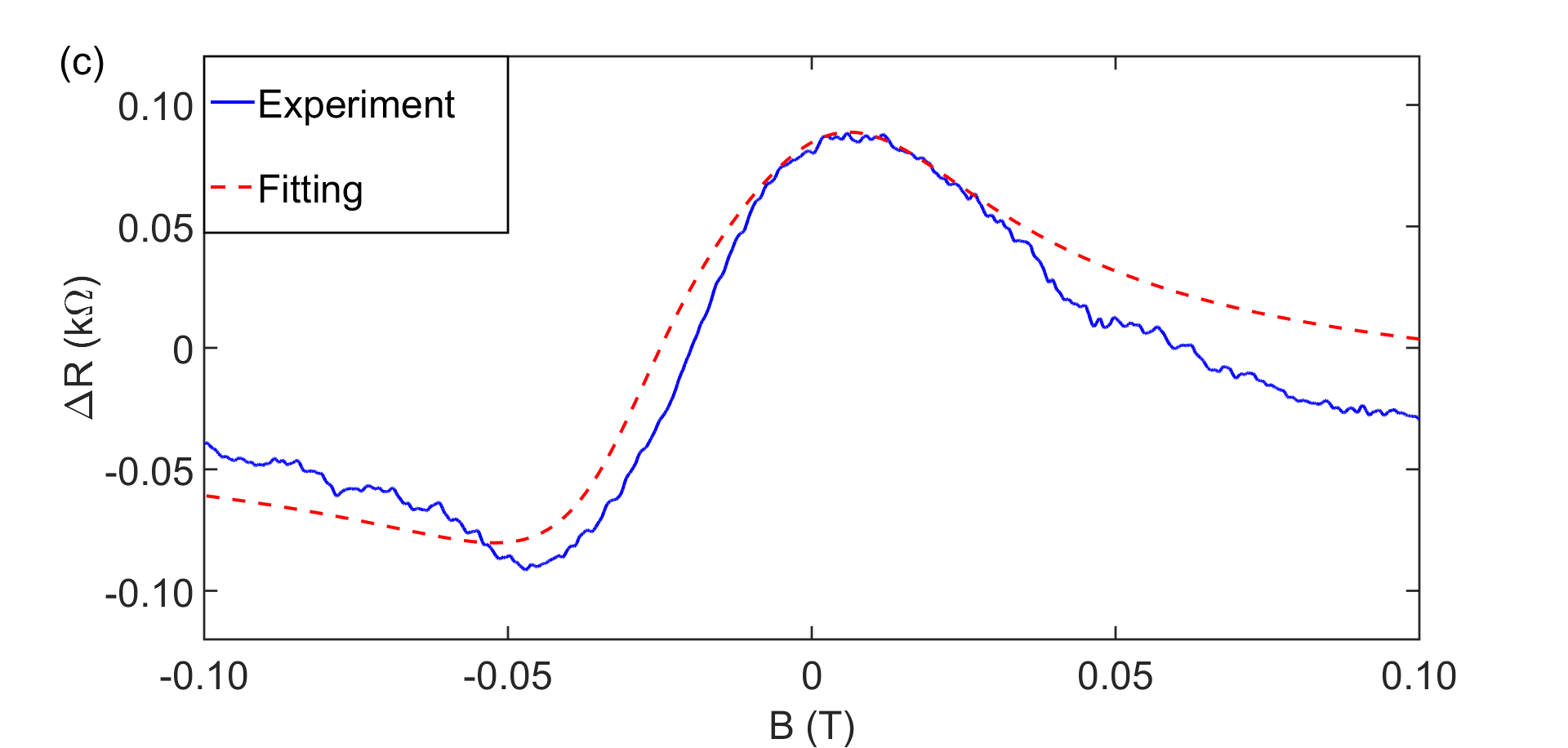}
		\label{fig:B_Fano}
	}%

	\caption{Magnetoresistance of the arch-QPC-reflector assembly. (a) The reflector is grounded or set to -0.25 V; the results show a typical Hall voltage development as the magnetic field is swept for different QPC voltage. Data for V$_r$ = - 0.25 V have been offset vertically by 50 $\Omega$ for clarity. (b) The reflector is set to - 0.3 V, QPC voltage V$_a$ is set at - 0.63 (at fine oscillations), - 0.24 (peak I), - 0.21 (peak II) and 0 V, respectively. Data have been offset vertically by 100 $\Omega$ for clarity. (c) Theoretical fitting of $\Delta$R(V$_a$,B) = R(V$_a$,B) - R(0,B) where V$_a$ = -0.24 V, showing the experimental data follow a Fano line shape, the Fano factor \textit{q} = 0.823.
	}           
	\label{fig:magnetoresistance}
\end{figure} 

The localized states, which represent the quantization of standing waves in the cavity, is highly dependent on the trajectory of electrons and thus sensitive to the perpendicular magnetic field. As a consequence, the fine oscillations arising from coupling between the cavity and 1D states should also be field sensitive\cite{MTM94}. In Fig.~\ref{fig:field_raw_data}, we show (negative) transverse magnetic field dependence of the resistance, which gives information on the coupling between 1D states and cavity states. It is clear that the fine oscillations smear out at a weak field of -60 mT as expected. The double-peak structure (in 1D-2D transition regime) follows a more complicated trend. The intensity of peak I gets reduced by increasing field but it survives at much higher field compared to the fine oscillations (- 120 mT at negative field end), on the contrary, peak II gets enhanced by the magnetic field\cite{CSM17}. Similar results were obtained with positive transverse magnetic field. We note that when peaks I and II are observed, the channel is much wider than that for the fine structure. Consequently the restriction on the formation of the localized state is much weaker accounting for the higher field necessary to remove the peaks. The initial enhancement of peak II is presumably a geometric effect.

A detailed magnetoresistance study is presented in Fig.~\ref{fig:magnetoresistance}. When the reflector is grounded or set to -0.25 V, a typical Hall voltage development is seen for both V$_a$ = 0 V where QPC is in 2D regime and V$_a$ = - 0.5 V when the quasi-1D channel forms in the QPC, as shown in Fig.~\ref{fig:magnetoresistance}(a). In Fig.~\ref{fig:magnetoresistance}(b), a voltage of - 0.3 V is applied to the reflector. When V$_a$ is 0 and - 0.21 V (cavity is not formed) the result is almost similar to that in Fig.~\ref{fig:magnetoresistance}(a),  however, SdH oscillations are seen in the large field regime which come from the contribution of the reflected electrons. When the cavity is switched on (V$_a$ = - 0.24 V or - 0.63 V), additional structure which is highly asymmetric against magnetic field is observed in the small field regime up to $\pm$ 70 mT, which is similar to the magnetic field value required to quench the fine oscillations as shown in Fig.~\ref{fig:field_raw_data}. In the large field regime a superposition of Hall voltage and SdH oscillations dominates. The dynamics in resistance with cavity switched on, e.g. green trace in Fig.~\ref{fig:magnetoresistance}(b), defined as $\Delta$R(V$_a$,B) = R (V$_a$, B) - R(0, B), also resembles a Fano resonance. Figure~\ref{fig:magnetoresistance}(c) shows a theoretical fitting of $\Delta$R(- 0.24 V, B) (corresponding to green trace in Fig.~\ref{fig:magnetoresistance}(b)) with the Fano line shape. It was also noticed that there was a shift of minima of \textit{R} against magnetic field in Fig.~\ref{fig:magnetoresistance}, which is likely due to the combined effect of inclined reflector and the negative field  because both of them guide electrons to Ohmic 3 while positive field directs electrons to Ohmic 4  whose trajectory is then compensated by the reflector. The difference becomes more apparent when the electrons are more ballistic and collimated (i.e. with more negative V$_a$). 
 
The self-consistency between the reflector [Fig.~\ref{fig:reflector_sweep}(b)] and magnetic field [Fig.~\ref{fig:magnetoresistance}(c)] induced Fano resonance shows that we can tune coupling between the 1D and cavity states both electrostatically and magnetically. To gain a comprehensive understanding of the scenario, a non-Hermitian description of the QPC-cavity hybrid system comprising of reflected wave functions into the continuum is necessary\cite{HI17}. This is beyond the scope of current work.

\begin{figure}
       
    \includegraphics[height=1.75in,width=3.5in]{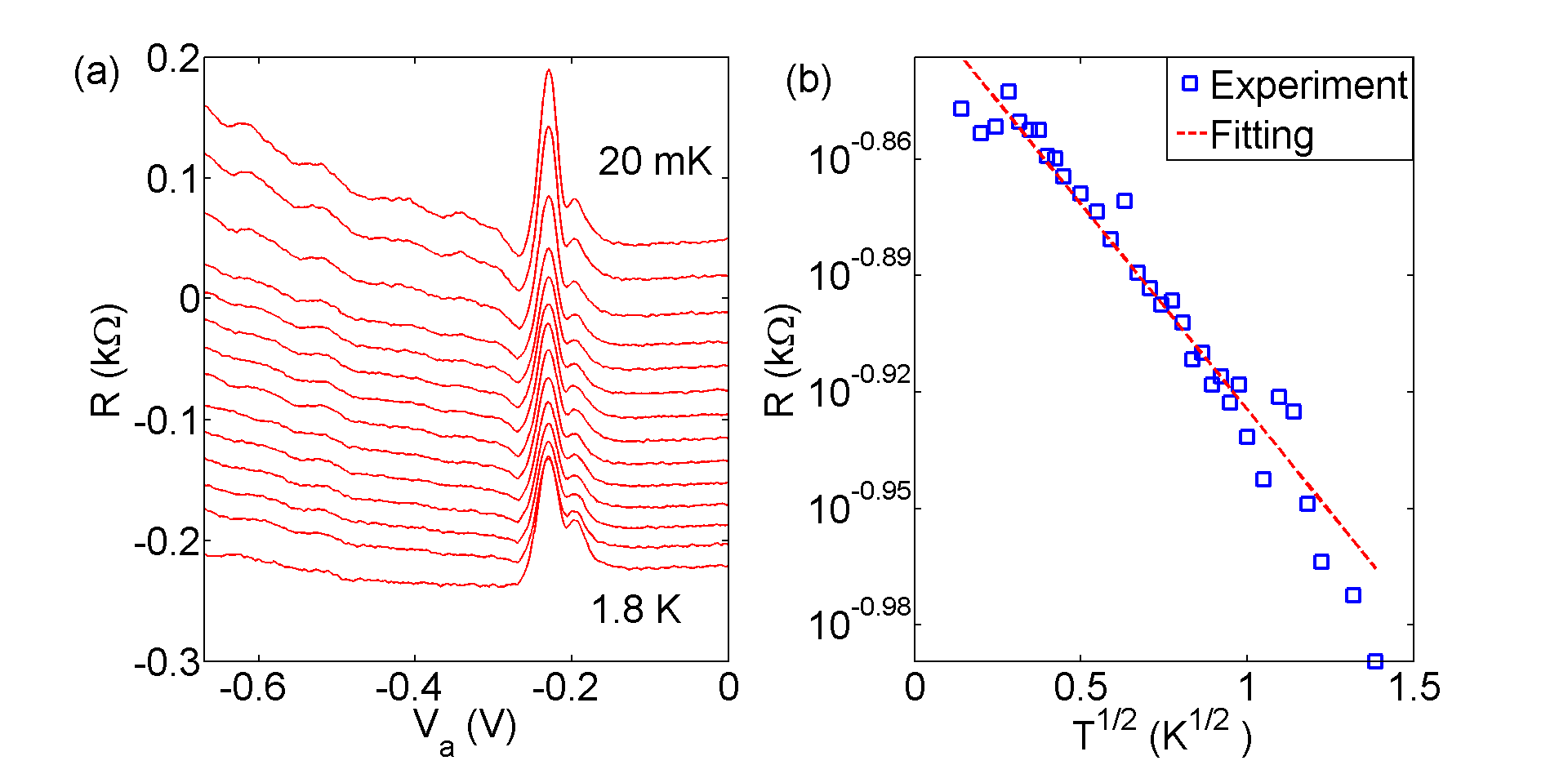}
 
	\caption{Temperature dependence of \textit{R}. (a) The evolution of non-local resistance at V$_r$= - 0.3 V against temperature; the lattice temperature increases from 20 mK (top trace) to 1.8 K (bottom trace). Data have been offset vertically for clarity. (b) Theoretical fitting of the height of the peak I of the double-peak structure using Eq.~(3) $\&$ (4) with \textit{p} = 1 for the fitting. 
	}           
	\label{fig:temperature_raw_data}
\end{figure} 

\subsection{Temperature dependence}

Temperature dependence measurement is a useful tool to investigate quantum effect such as the origin of fine oscillations. As mentioned previously, the energy spacing for cavity states is an order of magnitude smaller than the 1D subband spacing in the QPC, a slightly higher temperature makes the cavity states become a continuum while 1D subbands are still well resolved. The thermal smearing of the cavity states in turn  leads to smearing out of the fine oscillations. Figure~\ref{fig:temperature_raw_data}(a) shows the evolution of the fine oscillations in the lattice temperature range of 20 mK to 1.8 K. It is clear that the fine oscillations smear out at 1.8 K where the conductance plateaus of QPC still persist. On the other hand, the  double-peak structure is present even at 1.8 K; however, its intensity decreases against rising temperature. The double-peak structure is a consequence of interference effect, and its intensity can be expressed by\cite{YUC91},
$$ R \propto exp(-l/l_{\phi}) \eqno(3) $$ where \textit{l} is the electron propagation length and the temperature dependence of  phase coherence length $l_\phi$ is,
$$l_\phi \propto \sqrt{  T^{-p} } \eqno(4)$$
where $p=1$ for a 2D system\cite{MRM81} and $p=\frac{2}{3}$ for a 1D system\cite{PPG88}. In our device $p=1$ gives the best fitting as shown in Fig.~\ref{fig:temperature_raw_data}(b) which in consistent with the fact that the cavity states are essential to observe the  quantum interference\cite{note1}.

\section{Conclusion}

In conclusions, we have shown the operation of an integrated quantum device consisting of an arch-shaped QPC coupled to an electronic cavity, whose states can be tuned using a reflector gate. We have demonstrated that it is possible to couple and decouple the 1D states with the 2D cavity states, using either the reflector barrier or a transverse magnetic field, resulting in the direct observation of  Fano resonance which arises from the  interference between the QPC (1D) and cavity states (2D). Present results show the promise of such integrated quantum devices in realizing complex quantum systems to study the 1D-2D transition and a possible precursor for quantum information processing based on modulation of 1D states, for instance a quantum analogy of amplitude-modulation (AM) or phase-modulation (PM) in classical in classical information processing.

The work is funded by the Engineering and Physical Sciences Research Council (EPSRC, UK).


\begin{thebibliography}{10}
\bibitem{note1}
\bibinfo{year}{The closed-square cavity formed near Ohmic 5 does not affect the observations. The dimension of the square cavity is larger than the phase coherence length extracted from the temperature data. Ohmic 6 is floating all the time. }
\end{thebibliography}
\end{document}